\documentclass[letterpaper]{jpconf}
\usepackage{graphicx}

\begin{document}
\title{High {\boldmath $Q^{2}$} Measurements from HERA}

\author{David M. South, on behalf of the H1 and ZEUS Collaborations}

\address{Technische Universit\"{a}t Dortmund, Experimentelle Physik V, 44221 Dortmund, Germany}

\ead{david.south@desy.de}

\begin{abstract}
New measurements of neutral and charged current cross sections at large
negative four--momentum transfer squared $Q^{2}$ have been performed by
H1 and ZEUS, using up to the complete HERA~II $e^{\pm}p$ data, which was taken
with polarised electron and positron beams.
The data are compared to predictions of the Standard Model, based on various parton
distribution function parameterisations.
The polarisation asymmetry of the neutral current interaction is measured as a function of
$Q^{2}$, as well as the polarisation dependence of the charged current cross section
and both are found to be in agreement with the Standard Model expectation.
The HERA~II cross sections are also combined with previously published HERA~I
data to obtain the most precise unpolarised measurements.

\end{abstract}

\section{Introduction}

HERA measurements of proton structure in neutral current (NC) and
charged current (CC) deep inelastic scattering (DIS) with polarised
lepton beams are crucial to the understanding of the detailed dynamics
of QCD as well as allowing the chiral structure of electroweak
interactions to be simultaneously probed at the highest energies.
The NC interaction $ep \rightarrow ep$ proceeds via $\gamma$ or $Z^{0}$
exchange, whereas the CC interaction $ep \rightarrow e\nu$ involves
the exchange of a $W^{\pm}$ boson.
The cross sections of such processes are described in terms of the negative
four--momentum transfer squared, $Q^{2}$, the fraction of proton momentum
carried by the struck quark, Bjorken $x$, and the inelasticity of the interaction,
$y$, which are related according to the expression $Q^{2} = sxy$. 
The first measurements of such processes to include the full HERA~II ($2003$--$2007$)
data are now available, and a selection of results from the H1~\cite{h1nc09,h1cc09}
and ZEUS~\cite{zeusncEplus,zeusncEminus,zeusccEminus,zeusccEplus}
Collaborations is presented here.


The HERA~II data were taken with a polarised incident electron or positron
beam of energy $27.6$~GeV in collision with an unpolarised proton
beam of energy $920$~GeV, yielding a centre-of-mass
energy of $\sqrt{s}=319$~GeV.
The $e^{+}p$ and $e^{-}p$ data sets are further subdivided into periods of
positive and negative longitudinal polarisation, $P_e=(N_R-N_L)/(N_R+N_L)$,
where $N_R$ ($N_L$) is the number of right (left) handed leptons in the beam.
The typical lepton beam polarisation of the data is around $30$--$40$\%.
Unpolarised measurements are performed by correcting for the residual beam
polarisation according to the SM prediction, where H1 measurements~\cite{h1nc09,h1cc09}
are also combined with the HERA~I ($1994$--$2000$) data.

\section{Unpolarised measurements}

\begin{figure}[h]
\begin{minipage}{20pc}
\includegraphics[width=20pc]{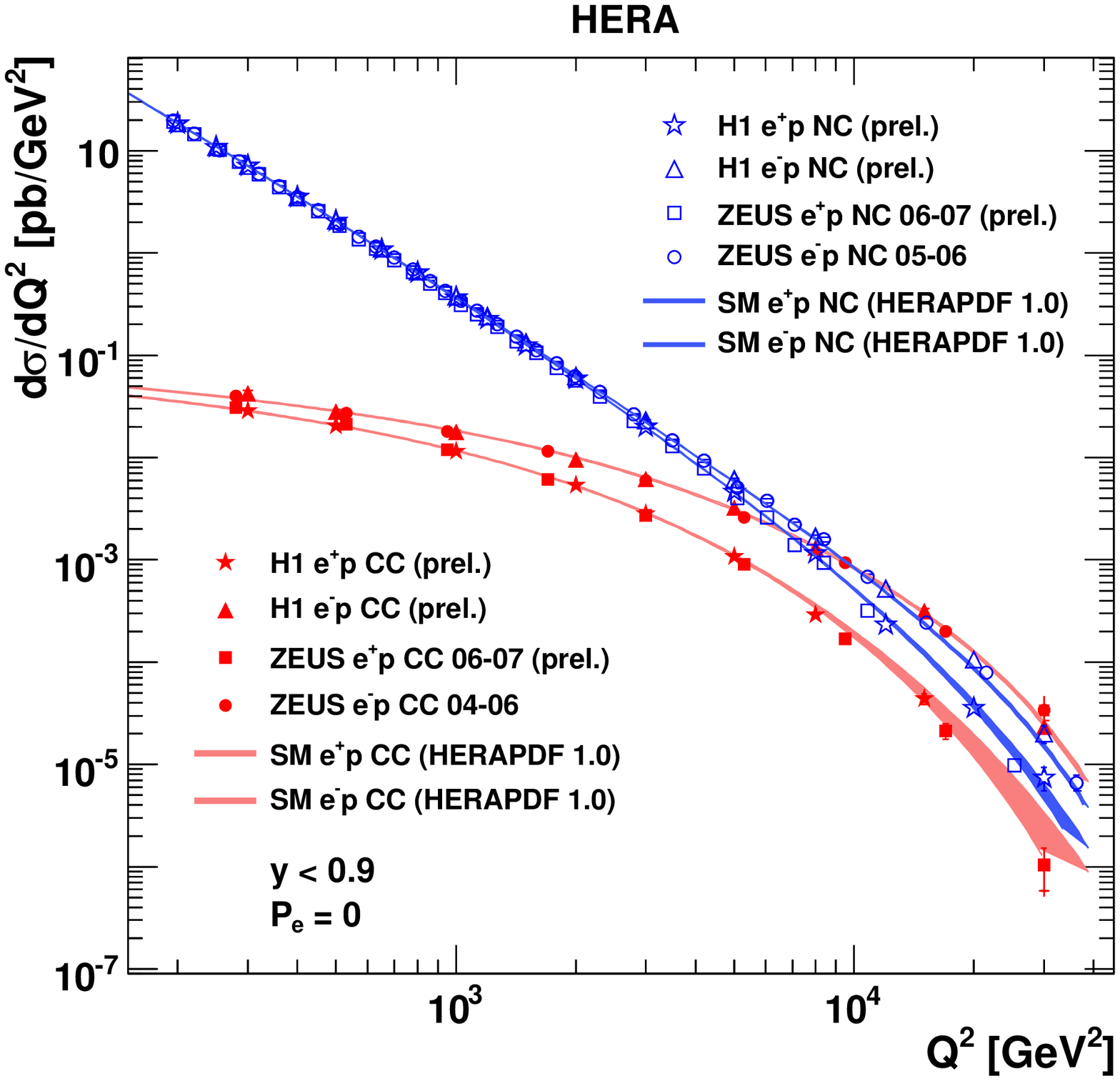}
\end{minipage}
\begin{minipage}{18pc}
\includegraphics[width=18pc]{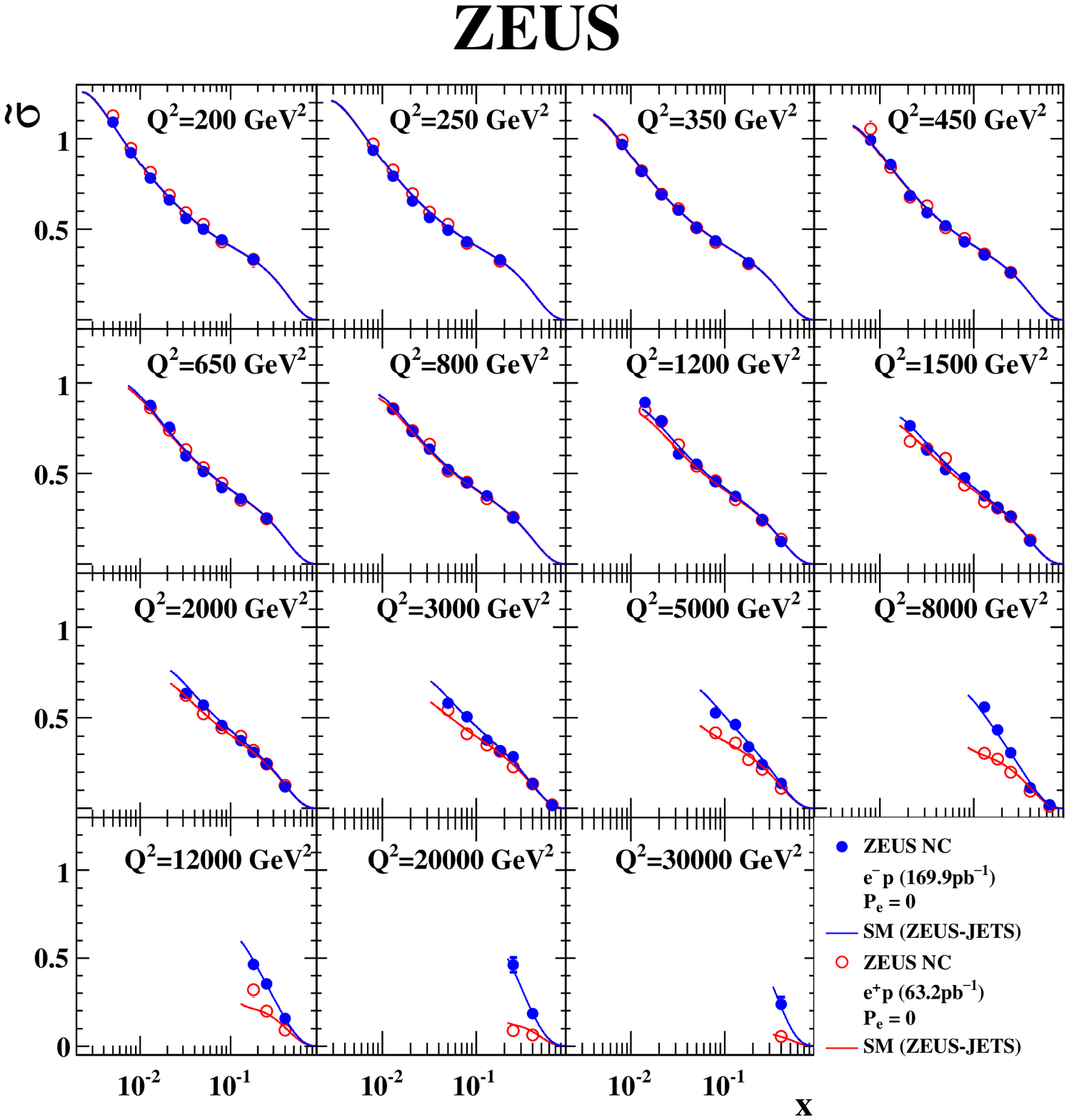}
\end{minipage} 
\begin{minipage}{19pc}
  \caption{\label{fig1} The inclusive unpolarised NC~(open symbols) and CC~(solid symbols) DIS
    cross sections measured by H1 and ZEUS as a function of $Q^2$ using the
    $e^{+}p$~(stars, squares) and $e^{-}p$~(triangles, circles) HERA data.
    The inner and outer error bars represent respectively the statistical and total errors.
    The data are compared to the SM prediction based on the HERAPDF $1.0$ parameterisation (shaded band).
  }
\end{minipage}\hspace{2pc}%
\begin{minipage}{17pc}
  \caption{\label{fig2}The reduced NC cross section $\tilde{\sigma}(x,Q^{2})$ measured by ZEUS as a function
    of $x$ in different bins of $Q^2$ for unpolarised $e^{+}p$ (open points) and $e^{-}p$ (solid points) data.
    The inner and outer error bars represent respectively the statistical and total errors.
    The data are compared to the SM prediction based on the ZEUS--JETS PDF parameterisation (lines).
  }
\end{minipage} 
\end{figure}

The $Q^{2}$ dependence of the unpolarised NC and CC cross sections at HERA is shown
in figure~\ref{fig1}.
The H1 and ZEUS $e^{\pm}p$ data are well described by the SM prediction, which is based
on the HERAPDF $1.0$ parameterisation of the parton densities in the proton~\cite{herapdf}.
It can be seen that at low $Q^{2}$ the cross sections differ significantly, due to the difference
in the propagator terms in the cross section: for NC the photon exchange goes as
$\sim 1/Q^{4}$, whereas the CC process includes the mass of the $W$ boson and is
proportional to $[M_{W}^{2}/(M_{W}^{2}+Q^2)]^{2}$.
The NC and CC cross sections are of comparable size for $Q^{2} \approx M_{W}^{2}$
and higher, illustrating the unification of the electromagnetic and the weak interactions.


The difference between the $e^{+}p$ and $e^{-}p$ CC cross sections arises from the
less favourable helicity factor of $(1-y)^{2}$ in the case of  $e^{+}p$ collisions and 
the difference between the up and down quark distributions in the proton.
For the NC process, the $e^{-}p$ cross section is larger than in $e^{+}p$ collisions
for values of $Q^{2} \approx M_{Z^{0}}^{2}$ and higher, due to the interference between the pure
$\gamma$ and $Z^{0}$ exchange, which is positive in the case of  $e^{-}p$ scattering.
This can be seen in more detail in the reduced unpolarised double differential NC cross
section ${\tilde\sigma}(x,Q^{2})$, as shown measured by ZEUS in figure~\ref{fig2},
where the separation of the $e^{+}p$ and $e^{-}p$ measurements at
low $x$ and high $Q^{2}$ is clearly visible, in agreement with the SM prediction from
the ZEUS--JETS PDF.

\section{Measurements using polarised data}

\begin{figure}[h]
\begin{minipage}{19pc}
\includegraphics[width=19pc]{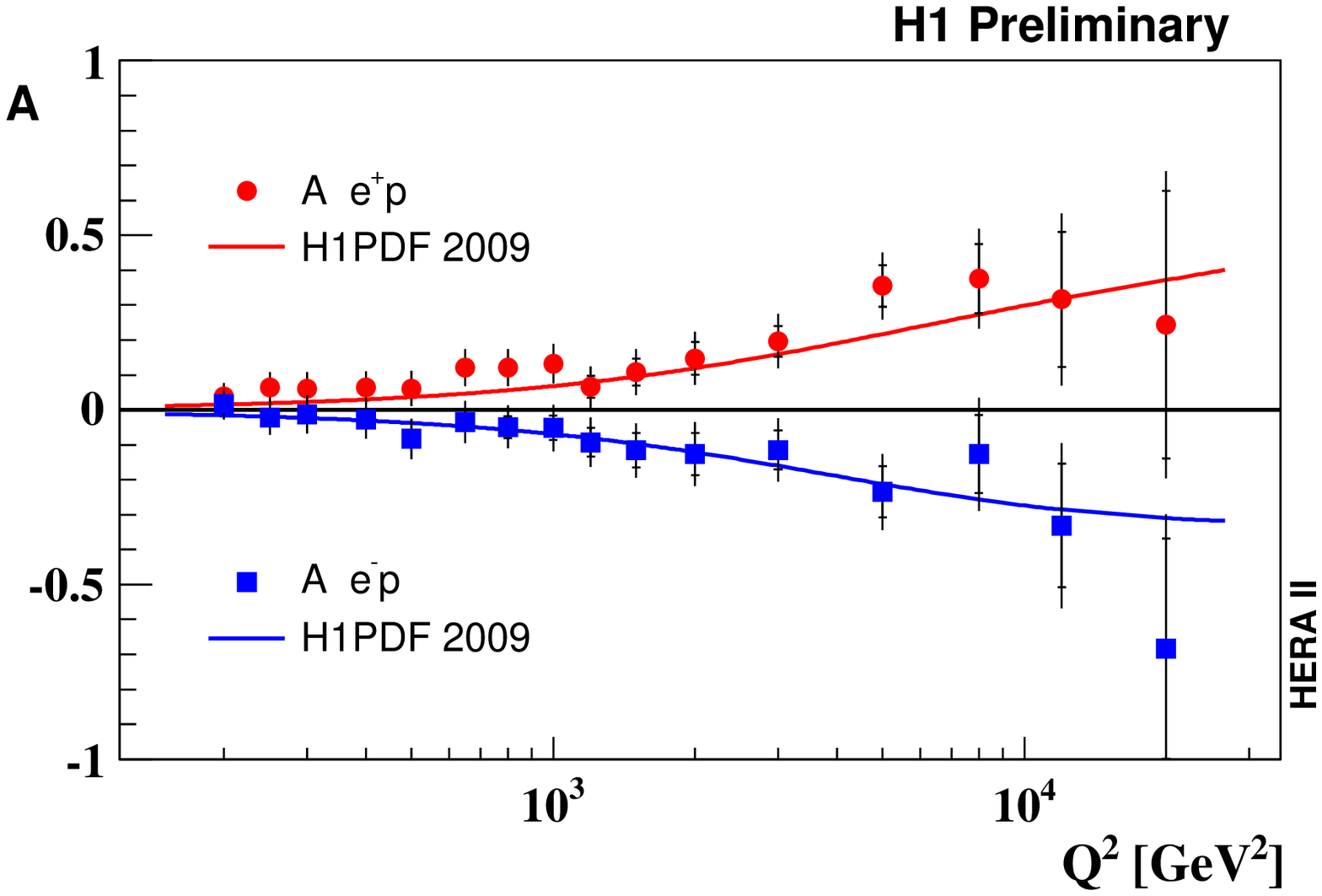}
\end{minipage}
\begin{minipage}{19pc}
\includegraphics[width=19pc]{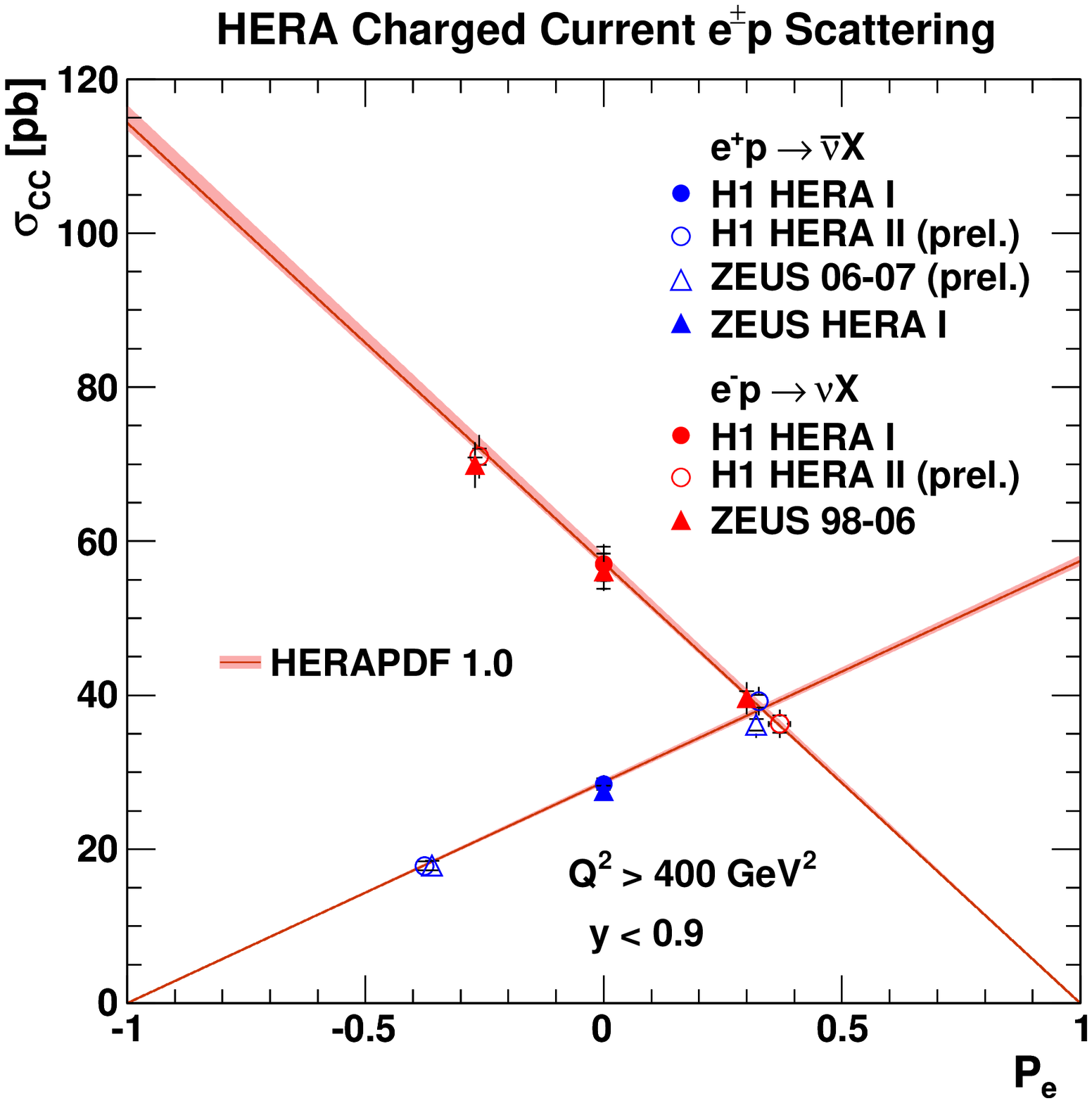}
\end{minipage} 
\begin{minipage}{18pc}
  \caption{\label{fig3}The $Q^2$ dependence of the polarisation asymmetry, $A$, measured
    by H1 using the $e^{+}p$~(circles) and $e^{-}p$~(squares) NC DIS data.
    The inner and outer error bars represent respectively the statistical and total errors.
    The data are compared to the SM prediction based on the H1PDF $2009$ parameterisation (lines).

  }
\end{minipage}\hspace{2pc}%
\begin{minipage}{18pc}
  \caption{\label{fig4}The dependence of the integrated $e^{\pm}p$ CC DIS cross section
    $\sigma_{\rm CC}$ on the lepton beam polarisation $P_{e}$ measured by H1 and ZEUS.
    The inner and outer error bars represent respectively the statistical and total errors.  The data
    are compared to the SM prediction based on the HERAPDF $1.0$ parameterisation (shaded bands).
 }
\end{minipage} 
\end{figure}

The Standard Model predicts a difference in the NC cross section
for leptons with different helicity states arising from the chiral
structure of the neutral electroweak exchange.
With longitudinally polarised lepton beams in HERA~II such polarisation
effects can be tested, providing a direct measure of the electroweak effects
in the neutral current cross sections.
In effect, there are then four HERA~II data sets to consider: electron and positron
collisions, with left-- or right--handed polarised leptons. 
The polarisation asymmetry is calculated as:
$A^{\pm} = 2 / (P_{e, +}-P_{e,-}) \cdot [\sigma^{\pm}(P_{e,+}) -\sigma^{\pm}(P_{e,-})] / [\sigma^{\pm}(P_{e,+}) +\sigma^{\pm}(P_{e,-})]$,
where, for example $\sigma^{+}(P_{e,+})$ is the measured cross section for the positively polarised $e^{+}p$ data set.
In positron scattering $A$ is expected to be positive and about equal
to $-A$ in electron scattering.
The asymmetry measurement performed by H1 as a function of $Q^{2}$ is shown
in figure~\ref{fig3}, compared to the SM expectation based on the H1PDF $2009$ parameterisation.
The magnitude of the asymmetry is observed to increase with increasing $Q^{2}$
and is negative in $e^{-}p$ and positive in $e^{+}p$ scattering, in agreement with the
SM prediction and confirming parity violation in neutral current interactions.

\begin{figure}[t]
\begin{minipage}{18.5pc}
\includegraphics[width=18.5pc]{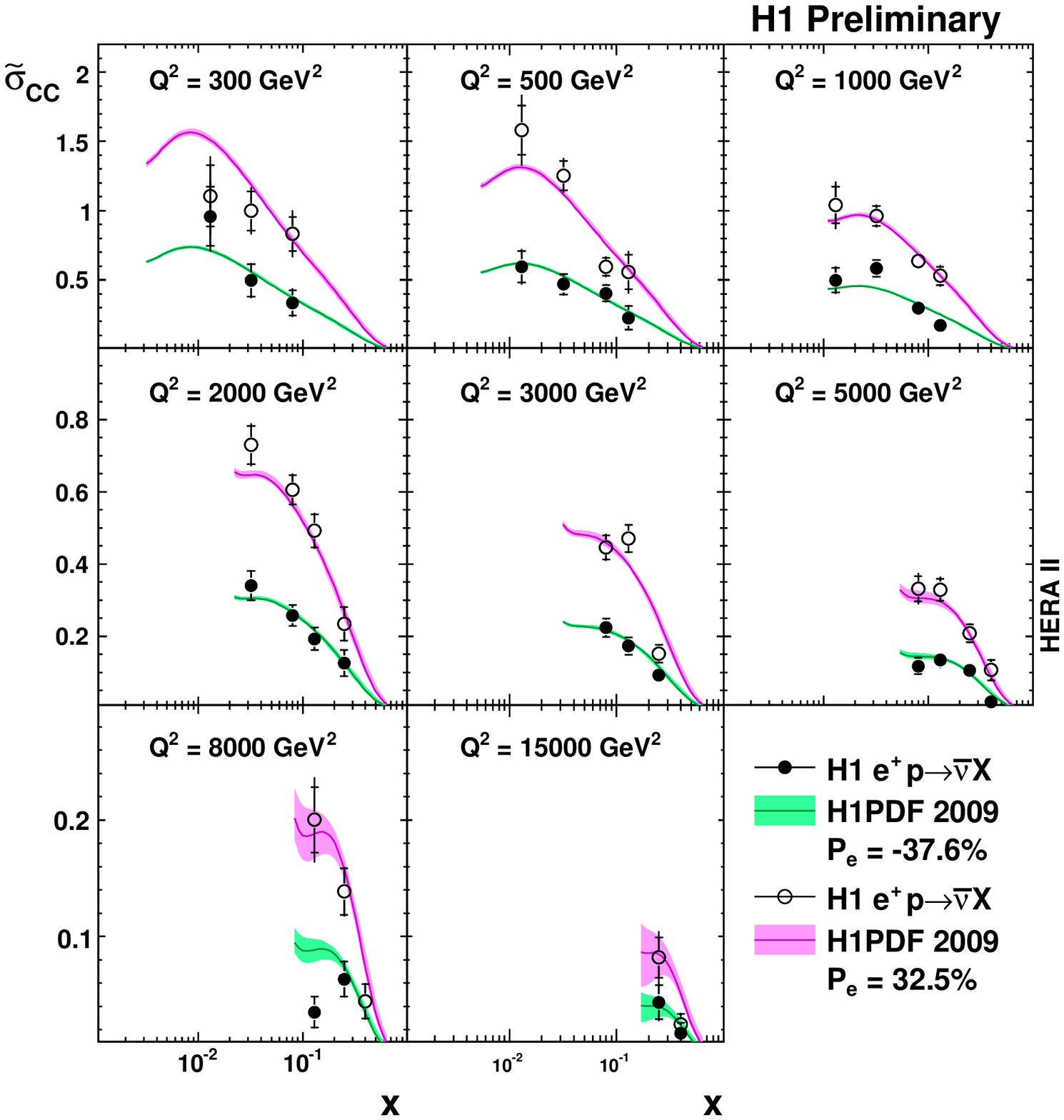}
\end{minipage}
\hspace{1pc}
\begin{minipage}{18.5pc}
\includegraphics[width=18.5pc]{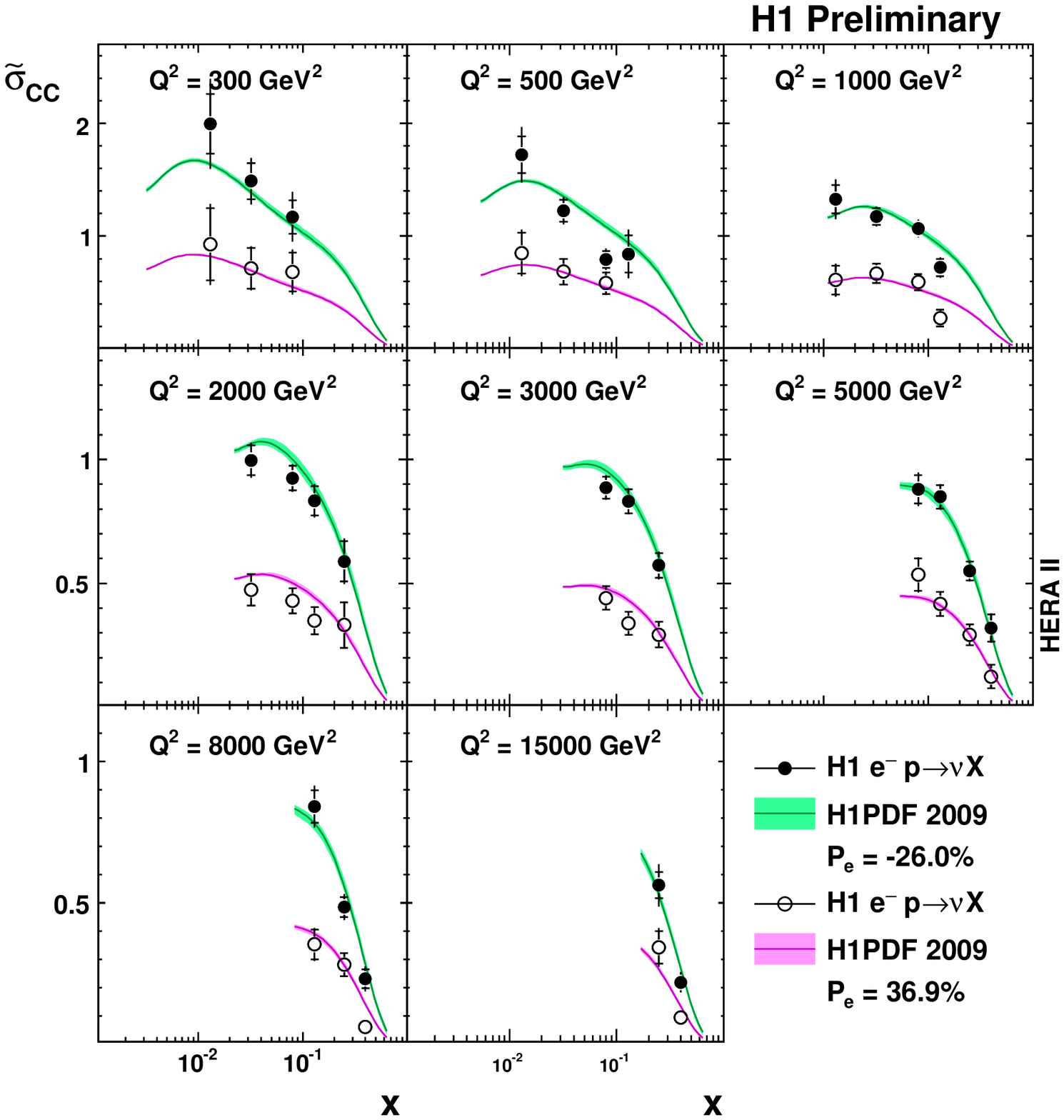}
\end{minipage} 
\begin{minipage}{38pc}
\caption{\label{fig5} The polarised reduced CC DIS cross section $\tilde{\sigma}_{\rm CC}(x,Q^{2})$ measured by H1 using
  $e^+p$~(left) and $e^-p$~(right) data as a function of $x$ in different bins of $Q^2$.
  The negatively (positively) polarised data are shown by the and solid (open) points.
  The inner and outer error bars represent respectively the statistical and total errors.
  The data are compared to the SM prediction based on the H1PDF $2009$ parameterisation (shaded bands).
  }
\end{minipage} 
\end{figure}

The SM predicts the absence of right handed charged currents and a
linear dependence of the CC cross section on the lepton beam polarisation,
which is given by the relationship: $\sigma_{CC}^{\pm}(P_{e}) = (1 \pm P_{e}) \cdot \sigma_{CC}^{\pm}(P_{e}=0)$.
In other words, for a fully right--handed electron beam ($P_e=1$) or a fully
left--handed positron beam ($P_e=-1$) the SM cross section is identically
zero and any deviation from this linear dependence would indicate new physics.
The integrated CC cross section measured by H1 and ZEUS in the kinematic
region $Q^{2}> 400$~GeV$^{2}$ and $y<0.9$ is shown in figure~\ref{fig4}.
Measurements of the unpolarised total cross section in the same phase
space based on HERA~I data are also shown.
The measurements are compared to the SM expectation based on
the HERAPDF $1.0$ parameterisation.
The data exhibit a clear linear polarisation dependence of the cross section, which
is maximal for left--handed $e^{-}p$ scattering and right--handed $e^{+}p$ scattering
in agreement with the SM prediction and demonstrating the parity violation of purely
weak charged current interactions. 
The polarised reduced double differential CC cross section ${\tilde\sigma_{CC}}(x,Q^{2})$
is shown in figure~\ref{fig5} as measured by H1 using the HERA~II $e^{+}p$ (left) and 
$e^{-}p$ (right) data.
The polarisation asymmetry is again clearly visible where the positively (negatively) polarised CC
cross section is higher for $e^{+}p$ ($e^{-}p$), in agreement with the prediction from H1PDF $2009$.

\section{Conclusions}

Measurements by H1 and ZEUS of polarised and unpolarised neutral and charged current cross
sections at HERA are presented.
The data are found to be consistent with the predicted behaviour of
polarised $ep$ scattering in the Standard Model. 
The high $Q^{2}$ data from H1 and ZEUS, separately and ultimately in combination,
will provide more constraints on the structure of the proton and input into electroweak fits of
the HERA data and QCD fits such as HERAPDF, which are a vital ingredient to the success of
proton colliders like the LHC.

\end{document}